\documentclass{article}

\usepackage{microtype}
\usepackage{graphicx}
\usepackage{subfigure}
\usepackage{booktabs} 

\usepackage{hyperref}

\usepackage[accepted]{icml2023}

\usepackage{amsmath}
\usepackage{amssymb}
\usepackage{mathtools}
\usepackage{amsthm}

\usepackage[capitalize,noabbrev]{cleveref}

\theoremstyle{plain}

\theoremstyle{definition}

\theoremstyle{remark}

\usepackage[textsize=tiny]{todonotes}

\icmltitlerunning{Field-Level Inference with Microcanonical Langevin Monte Carlo}

\begin{document}

\twocolumn[
\icmltitle{Field-Level Inference with Microcanonical Langevin Monte Carlo}



\icmlsetsymbol{equal}{*}

\begin{icmlauthorlist}
\icmlauthor{Adrian E.~Bayer}{princeton,cca,bccp,lbl,equal}
\icmlauthor{Uro\v{s} Seljak}{bccp,lbl,equal}
\icmlauthor{Chirag Modi}{cca,ccm}
\end{icmlauthorlist}

\icmlaffiliation{princeton}{Department of Astrophysical Sciences, Princeton University, Princeton, NJ, USA}
\icmlaffiliation{cca}{Center for Computational Astrophysics, Flatiron Institute, New York, NY, USA}
\icmlaffiliation{ccm}{Center for Computational Mathematics, Flatiron Institute, New York, NY, USA}
\icmlaffiliation{bccp}{Department of Physics, University of California, Berkeley, CA, USA}
\icmlaffiliation{lbl}{Lawrence Berkeley National Laboratory, Berkeley, CA, USA}

\icmlcorrespondingauthor{Adrian E.~Bayer}{abayer@princeton.edu}

\icmlkeywords{Machine Learning, ICML}

\vskip 0.3in
]



\printAffiliationsAndNotice{\icmlEqualContribution} 

\begin{abstract}
Field-level inference provides a means to optimally extract information from upcoming cosmological surveys, but requires efficient sampling of a high-dimensional parameter space.
This work applies Microcanonical Langevin Monte Carlo (MCLMC) to sample the initial conditions of the Universe, as well as the cosmological parameters $\sigma_8$ and $\Omega_m$, from simulations of cosmic structure.
MCLMC is shown to be over an order of magnitude more efficient than traditional Hamiltonian Monte Carlo (HMC) for a $\sim 2.6 \times 10^5$ dimensional problem. Moreover, the efficiency of MCLMC compared to HMC greatly increases as the dimensionality increases, suggesting gains of many orders of magnitude for the dimensionalities required by upcoming cosmological surveys.
\end{abstract}

\section{Introduction}
\label{sec:intro}

Modern cosmology has many open problems, ranging from measuring the neutrino mass to uncovering the nature of dark energy. 
To answer these questions, upcoming surveys of cosmic structure, such as 
DESI \citep{collaboration2016desi}, PFS \citep{Takada_2014},
Rubin Observatory LSST \cite{LSSTSci},
Euclid \cite{Euclid}, 
SPHEREx \cite{SphereX_2014}, 
SKA \cite{SKA_2009}, and Roman Space Telescope \cite{spergel2013widefield}
will provide unprecedented amounts of high-resolution data. These rich datasets promise much information from small scales, however, the nonlinear nature of these scales presents a challenge in optimally extracting the information. One promising method to extract all of the information is field-level inference, which performs Bayesian inference of the initial conditions for every voxel in the cosmic field, as well as the cosmological parameters. This involves sampling a high dimensional parameter space: for 
modern survey volumes of $10\,({\rm Gpc}/h)^3$ with resolutions of order $4\,{\rm Mpc}/h$
this leads to $10^8$
dimensions or more, which can be prohibitively slow to converge 
with 
standard samplers.

The traditionally used method for sampling in field-level inference is (Metropolis adjusted) Hamiltonian Monte Carlo (HMC) \cite{HMCDuane, HMCneal, conceptualHMC}. This method excels over pure Metropolis-Hastings (MH) by employing gradient information. However, in order to ensure detailed balance each step must undergo an accept-reject criterion, which in turn requires a small step size and reduces the efficiency. 
Recently, alternative sampling methods called Microcanonical Hamiltonian/Langevin Monte Carlo (MCHMC/MCLMC) \cite{robnik2022microcanonical,robnik2023microcanonical} have been introduced. These samplers
move fast in regions of 
low density and slow in the 
regions of high density, in 
contrast to HMC. They also do not require an accept-reject criterion, allowing a larger tolerance of energy fluctuations, in turn enabling a larger step size and larger effective sample size (ESS). Moreover, their ESS scales with dimensionality more favorably than HMC.

In this work we perform field-level inference to jointly sample the initial modes, and the cosmological parameters $\Omega_m$ and $\sigma_8$. The parameter $\Omega_m$ corresponds to the amount of matter in the Universe and $\sigma_8$ corresponds to the amplitude of density fluctuations.
There is an often reported tension between measurements of $\sigma_8$ made by different cosmological experiments \cite{Di_Valentino_2021}, thus using field-level inference to give an optimal measurement would aid in resolving this tension.

The structure of the paper is as follows. In §\ref{sec:background} we review field-level inference and outline the HMC, MCHMC, and MCLMC methods. In §\ref{sec:related} we discuss related work. In §\ref{sec:experiments} we apply the methods and analyse the ESS as a function of nonlinearity and dimensionality. 
We finally conclude in §\ref{sec:conclusions}.

\section{Background}
\label{sec:background}

We now outline field-level inference in §\ref{sec:field}, HMC in §\ref{sec:hmc}, MCHMC in §\ref{sec:mchmc}, and MCLMC in §\ref{sec:mclmc}.

\subsection{Field-Level Inference}
\label{sec:field}

The goal of field-level inference is to infer the initial modes of the Universe $s$ and/or the cosmological parameters $\lambda$. Given a forward model $f(s,\lambda)$, which in this work computes the 3d dark matter overdensity field, the posterior is given by
\begin{equation}
    -2\log P(s,\lambda|d) = \sum_{\vec{k}} \left[ \frac{|f_{\vec{k}}(s,\lambda) - d_{\vec{k}}|^2}{N_{\vec{k}}} + |s_{\vec{k}}|^2 \right],
\end{equation}
where the first term is the likelihood, with data $d$ and noise $N$, the second term is the prior, and the sum is over all voxels in $k$-space. Note that $s$ is defined to have unit variance.

Inferring $s$ corresponds to the task of reconstructing the initial conditions of the Universe, however, often $s$ are treated as nuisance parameters and one marginalizes over $s$ to yield only constraints on the cosmological parameters $\lambda$. 

\subsection{Hamiltonian Monte Carlo}
\label{sec:hmc}

The traditional approach for sampling in the context of field-level inference is HMC \cite{HMCDuane, HMCneal, conceptualHMC}. Given a $d$-dimensional target distribution $p(z) \propto e^{-\mathcal{L}(z)}$, where $z \in \mathbb{R}^d$, HMC uses the gradient $\nabla \mathcal{L}(z)$ to improve the sampling efficiency compared no-gradient methods such as MH. It considers the Hamiltonian $H(z,\Pi)$, where $\Pi$ is the canonical momentum, and samples the canonical ensemble in $2d$-dimension phase space, denoted by 
$p(z,\Pi) \propto e^{-H(z,\Pi)}$.
The success of HMC relies on the tuning of the Hamiltonian such that the marginal of $p(z,\Pi)$ over $\Pi$ converges to the target distribution, 
\begin{equation}
    p(z) \propto \int_{\mathbb{R}^d} d\Pi~ e^{-H(z,\Pi)}.
    \label{eqn:canonical}
\end{equation}
The most popular choice is the Hamiltonian of a particle in a potential, $H(z,\Pi)=\frac{1}{2}\Pi^2(z) + \mathcal{L}(z)$, for which the solution is the set of ODEs,
\begin{align}
    dz &= u dt,  \nonumber\\
    du &= -\nabla\mathcal{L}(z) dt,
    \label{eqn:eom_hmc}
\end{align}
where $t$ is time and $u$ is velocity.
Following Hamiltonian dynamics ensures the trajectory conserves the Hamiltonian, or energy, allowing efficient exploration at a fixed energy level. Different energy levels must be explored to obtain an accurate set of samples, which is achieved by resampling the momentum $\Pi$ according to its marginal distribution (a normal distribution) and results in inefficiencies \cite{conceptualHMC}.
Moreover, HMC additionally requires an MH accept-reject step, which necessitates a sufficiently small step size to ensure a frequent rate of acceptance. 


\subsection{Microcanonical Hamiltonian Monte Carlo}
\label{sec:mchmc}
Unlike HMC which considers the marginal of the canonical distribution, 
the approach of MCHMC is to tune the Hamiltonian such that the microcanonical distribution marginalized over the momentum variables gives the target distribution, as follows
\begin{equation}
    p(z) \propto \int_{\mathbb{R}^d} d\Pi~ \delta(H(z,\Pi)-E),
    \label{eqn:microcanonical}
\end{equation}
where $\delta(\cdot)$ denotes the delta function, and $E$ is the energy. 
There are many solutions for the Hamiltonian, but the one adopted for MCHMC yields the equations of motion,
\begin{align}
    dz &= u dt,  \nonumber\\
    du &= P(u) f(z) dt,
    \label{eqn:eom_mchmc}
\end{align}
where we have introduced the projection $P(u)\equiv(I-uu^T)$ and force $f(z) \equiv -\nabla \mathcal{L} (z)/(d-1)$ \cite{ESH, BIoptimization, robnik2022microcanonical}. The key difference to  HMC  in Eqn.~(\ref{eqn:eom_hmc}) is the projection. 
Unlike HMC, the MCHMC dynamics converges to the target distribution while maintaining a constant energy.

However, following Eqns.~\ref{eqn:eom_mchmc} exactly will not ensure ergodicity; for example, N\"{o}ther's Theorem implies that any symmetry will lead to a conserved quantity and in turn confine the dynamics to a subspace of the energy surface. 
Ergodicity can be achieved by energy-conserving momentum resampling \cite{billiards, BIoptimization, robnik2023microcanonical}, in which billiard-like bounces to the momentum are introduced such that the direction of the trajectory is changed, but the magnitude of the momentum, and thus the energy, is the same. Introducing momentum decoherence in this manner is analogous to the resampling of the momentum in HMC, but with fixed energy. The frequency of momentum resampling is a hyperparameter of MCHMC.

\subsection{Microcanonical Langevin Monte Carlo}
\label{sec:mclmc}

To further speed up reaching ergodicity, the ODEs can be modified by considering Langevin dynamics \cite{MALA,Girolami_2011} such that,
\begin{align}
    dz &= u dt, \\
    du &= P(u) \left[ f(z) dt + \eta dW \right],
    \label{eqn:eom_mclmc}
\end{align}
where $\eta$ is a hyperparameter and $W$ is a standard normal random vector. This additional term proportional to $\eta$ can be understood physically as a diffusion term which enforces better exploration of the target, in turn boosting ergodicity. 

MCLMC has two hyperparameters, the step size and the amount of 
noise $\eta$. Both of these parameters can be tuned during a burn-in stage by monitoring fluctuations in the energy and ensuring they are below a certain threshold. 

\begin{figure*}[ht!]
\begin{center}
\centerline{\includegraphics[width=2\columnwidth]{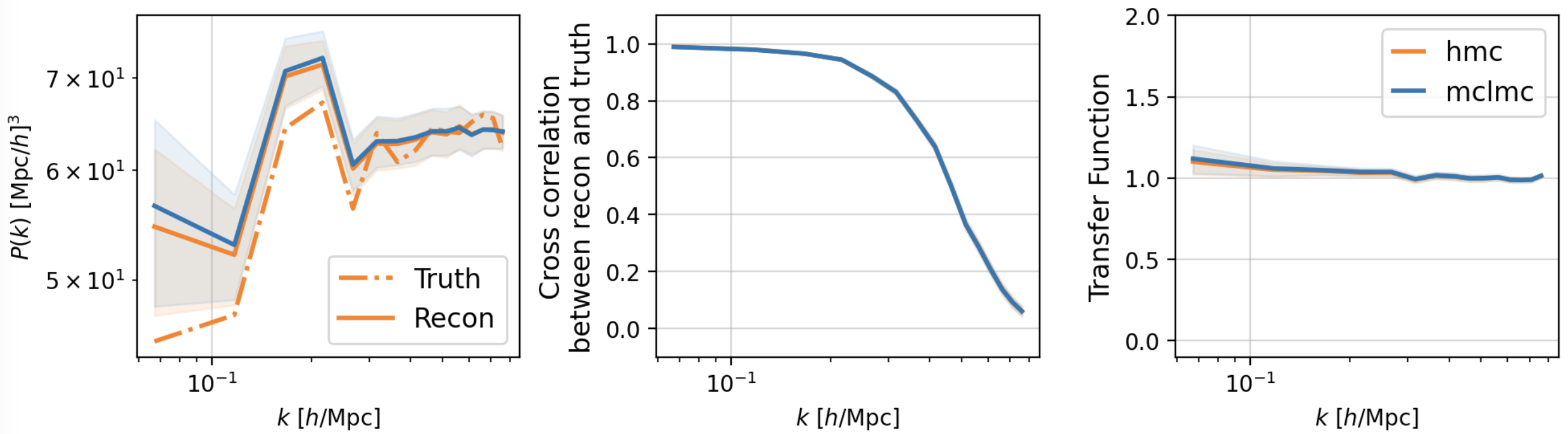}}
\caption{Samples of initial modes. Left: power spectrum of whitened modes, Center: cross-correlation, Right: transfer function.}
\label{fig:samples_z}
\end{center}
\vskip -0.2in
\end{figure*}

\begin{figure}[ht!]
\begin{center}
\centerline{\includegraphics[width=0.9\columnwidth]{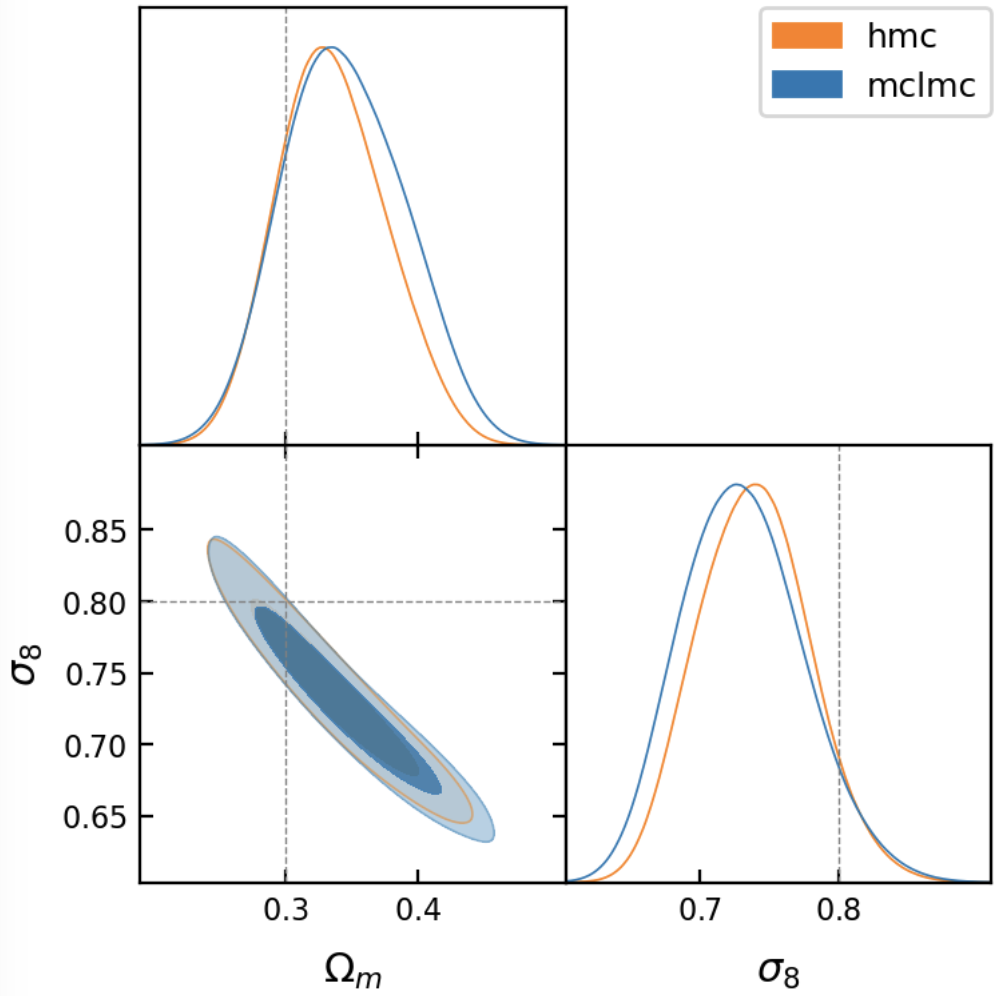}}
\caption{Samples of cosmological parameters. Dashed lines signify the truth.}
\label{fig:samples_q}
\end{center}
\vskip -0.2in
\end{figure}

\begin{figure}[ht!]
\begin{center}
\centerline{\includegraphics[width=\columnwidth]{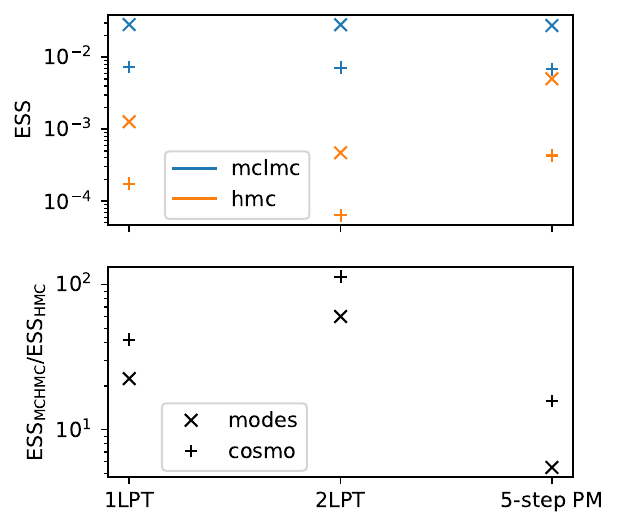}}
\caption{Comparison of ESS per gradient evaluation for different levels of nonlinearity. Top: ESS of MCLMC (blue) and HMC (orange), for the modes (x) and cosmological parameters (+). Bottom: Ratio of ESS from MCLMC versus HMC.}
\label{fig:ess_NL}
\end{center}
\vskip -0.2in
\end{figure}

\begin{figure}[ht!]
\begin{center}
\centerline{\includegraphics[width=\columnwidth]{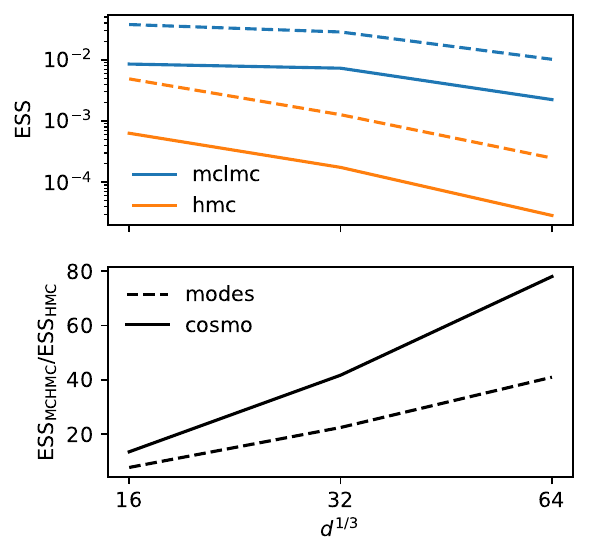}}
\caption{Scaling of ESS per gradient evaluation as a function of dimensionality $d$ for 1LPT. Top: ESS for MCLMC (blue) and HMC (orange), for the modes (dashed) and cosmological parameters (solid). Bottom: Ratio of ESS from MCLMC versus HMC.}
\label{fig:ess_d}
\end{center}
\vskip -0.2in
\end{figure}

\section{Related Work}
\label{sec:related}

Current approaches for sampling in field-level inference field-level inference include approximating the posterior around the optimum \cite{Seljak_2017, Millea_2022} and HMC \cite{Jasche_2013}. Hybrid HMC approaches based on variational inference have also been developed \citep{Modi:2022pzm}.

In terms of MCHMC, one type of microcanonical hamiltonian dynamics in the context of sampling was first explored by \citet{ESH}. 
The dynamics have also been applied in the context of optimization by \citet{BIoptimization}. A more general solution to sampling was provided by \citet{robnik2022microcanonical}, and the extension to Langevin dynamics was made by \citet{robnik2023microcanonical} in the context of lattice field theory. \citet{robnik2022microcanonical, robnik2023microcanonical} showed MCHMC/MCLMC to be superior to other HMC-based methods, such as NUTS \cite{NUTS}, which we will thus not consider here.

\section{Experiments}
\label{sec:experiments}

We simulate the nonlinear dark matter field using pmwd \cite{li2022pmwd}, a differentiable particle-mesh (PM) code written in JAX. It is capable of fitting $512^3$ particles on a single 80GB NVIDIA A100 GPU, and can evaluate one PM step in $\sim 0.1$ seconds. This code takes the cosmological parameters and the 3d intial gaussian field as input, and outputs the final 3d nonlinear dark matter field. It can perform nonlinear modeling to various levels of accuracy; we consider 3 levels of nonlineaarity, the Zel'dovich Approximation (ZA, or 1LPT), $2^{\rm nd}$ order Lagrangian perturbaiton theory (2LPT), and 2LPT followed by 5 steps of PM simulation. We consider three dimensionalities (corresponding to the number of voxels): $16^3, 32^3,$ and $64^3$. In each case the voxel resolution is $4\,{\rm Mpc}/h$. 

We perform inference over the initial modes in $k$-space (denoted $s$ in §\ref{sec:field}), and the cosmological parameters $\Omega_m$ and $\sigma_8$.
We precondition the modes by the theoretical posterior variance according to linear theory, i.e.~by applying the multiplicative factor $(P_L(k)+N)/N$, where $P_L$ is the linear power at the fiducial cosmology. This approximately ensures all voxels in $k$-space have unit variance, improving the sampler mixing. Similarly, we precondition the cosmological parameters to approximately have unit variance.

We run a chain of 10,000 samples with 2,000 steps of burn in (to be conservative). During HMC burn in, the step size is tuned to ensure a 65\% acceptance rate. We use 40 leapfrog steps for HMC, which we found to be optimal.
For MCLMC we tuned the step size and noise to ensure energy fluctuations per dimension are below a threshold of $10^{-4}$. We apply the Minimum-Norm integrator to solve the MCLMC ODEs.

Fig.~\ref{fig:samples_z} shows  the sampled modes for the $32^3$ 1LPT example. There is excellent agreement between HMC and MCLMC. Similarly, Fig.~\ref{fig:samples_q} shows the sampled cosmological parameters, again showing excellent agreement. This implies that MCLMC produces equally accurate samples to HMC.

Fig.~\ref{fig:ess_NL} compares the ESS per gradient evaluation between 1LPT, 2LPT, and 5-step PM. We consider the ESS for the modes and cosmological parameters separately. MCLMC outperforms HMC in all cases. The increase in ESS is greatest for 2LPT (two orders of magnitude), and lowest for 5-step PM. This is likely affected by the use of linear mode preconditioning, which we will improve upon in future work. 

In terms of the efficiency of the algorithms, Fig.~\ref{fig:ess_d} shows the ESS per gradient evaluation as a function of dimensionality for 1LPT. It can be seen in the top panel that while the ESS of HMC steadily decreases, the ESS of MCLMC decreases more mildly. This is because HMC requires a sufficiently small step size to prevent a relative energy error greater than unity: the energy scales with dimensionality, thus the step size required to achieve a relative error of less than unity decreases with dimensionality. On the other hand, MCLMC has no such constraint, meaning that for a fixed condition number there is no dependence of the ESS on dimensionality. In this case there is a slight increase in condition number as we include larger scale modes, meaning there is a slight decrease in ESS for MCLMC when moving to higher dimensions.
The lower panel of Fig.~\ref{fig:ess_d} shows how the ESS ratio of MCLMC versus HMC greatly increases with dimensionality. For $64^3$ the improvement is a factor of 40 for the modes and 80 for the cosmological parameter. This shows much promise to speed up sampling by many orders of magnitude when moving to the higher dimensionalities required to analyze upcoming cosmological data.

\section{Conclusions}
\label{sec:conclusions}

We have shown that MCLMC has over an order of magnitude higher ESS than HMC in the context of field-level inference. Moreover, the scaling of the ESS with dimensionality is more favorable for MCLMC, showing great promise to improve sampling efficiency by many orders of magnitude when moving to the higher dimensions required for upcoming cosmological data. We have also shown this for various levels of nonlinearity (1LPT, 2LPT, and 5-step PM); the improved efficiency of MCLMC makes it feasible to sample even when performing highly nonlinear forward modeling.

Future work includes application to the galaxy field, which will require sampling over additional parameters describing the matter--galaxy connection. Moreover, one can improve constraints by perform joint inference with other cosmological tracers, such as peculiar velocities \cite{Bayer:2022vid}.
Additionally, the preconditioning of the posterior could be improved to achieve a greater ESS, for example by fitting a normalizing flow to the posterior before running MCLMC \cite{hoffman2019neutralizing}.
Finally, the method will be applied to higher-dimensional maps which correspond to the data from upcoming cosmological surveys, including survey masks and systematic effects.

\if false

\begin{figure}[ht]
\vskip 0.2in
\begin{center}
\centerline{\includegraphics[width=\columnwidth]{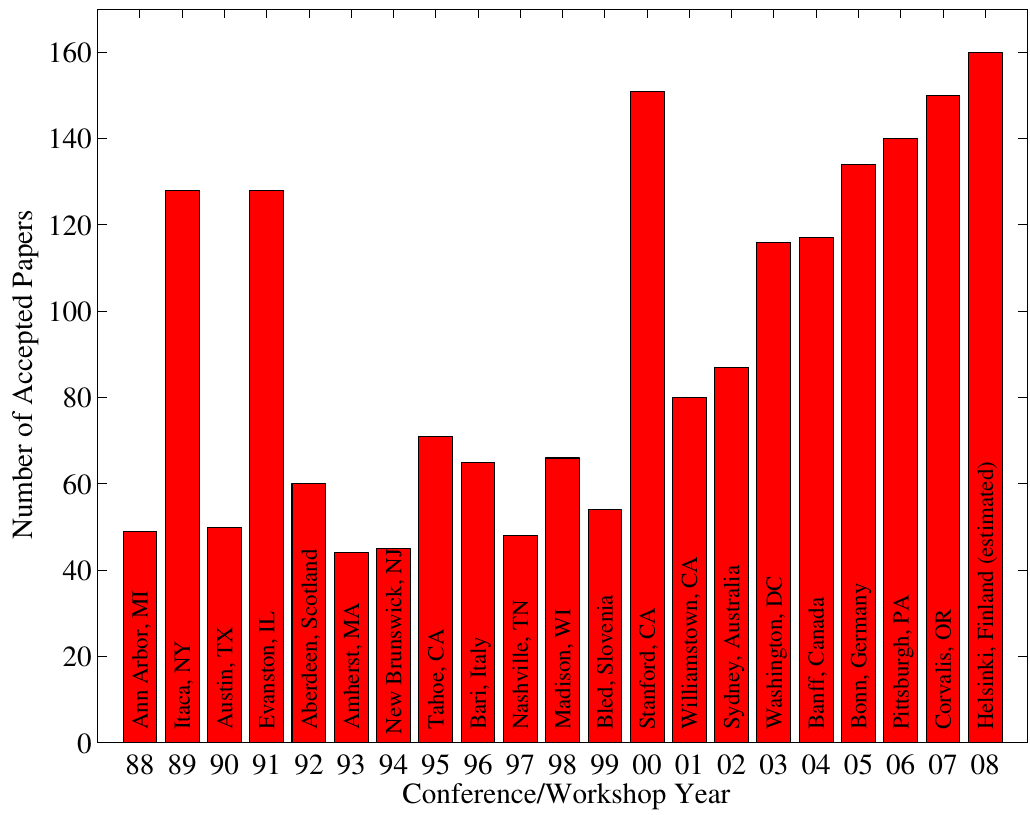}}
\caption{Historical locations and number of accepted papers for International
Machine Learning Conferences (ICML 1993 -- ICML 2008) and International
Workshops on Machine Learning (ML 1988 -- ML 1992). At the time this figure was
produced, the number of accepted papers for ICML 2008 was unknown and instead
estimated.}
\label{icml-historical}
\end{center}
\vskip -0.2in
\end{figure}

\begin{algorithm}[tb]
   \caption{Bubble Sort}
   \label{alg:example}
\begin{algorithmic}
   \STATE {\bfseries Input:} data $x_i$, size $m$
   \REPEAT
   \STATE Initialize $noChange = true$.
   \FOR{$i=1$ {\bfseries to} $m-1$}
   \IF{$x_i > x_{i+1}$}
   \STATE Swap $x_i$ and $x_{i+1}$
   \STATE $noChange = false$
   \ENDIF
   \ENDFOR
   \UNTIL{$noChange$ is $true$}
\end{algorithmic}
\end{algorithm}
\fi



\bibliography{example_paper}
\bibliographystyle{icml2023}



\end{document}